\documentstyle{article}
\begin{document}
\newcommand{\ex}{e^{-\tau}}
\newcommand{\ec}{e^{-\tau_{cr}}}
\newcommand{\fr}[2]{\frac{#1}{#2}}
\newcommand{\bern}{\frac{5}{2}T-A_3x+\frac{1}{2}\frac{x^4}{\rho^2}+
(1-e^{-\tau})A_1\frac{T^4}{\rho}}

\title{New approach to modeling of radiationaly driven envelopes
at arbitrary optical depths}

\section{Introduction}
Evolution of massive stars $(M>\sim 20 M)$ is accompanied by a mass loss,
initiated by a high luminosity and high radiation pressure. In blue
supergiants, situated near the main sequence, mass loss rate is moderate
$M\sim 10^{-6} M_{\bigodot}/yr.$ and is connected with the outflow of layers
having small optical depth. This mass loss rate is determined by radiation
pressure in lines, where spectral absorption coefficient may be very high.
Theory of the outflow is based hear on Sobolev approximation of the lines
treatment, ( CAK theory ), and velocity gradient in the flow increase
essentially the pushing effect of the radiation pressure (Castor, Abbot, Klein
, 1975). First
attempt to construct a theory of such mass outflow have been done by
~\cite{ls}, and for recent development of CAK theory see ~\cite{owo}.

Evolved massive stars may
lose mass with much higher rate then the blue supergiants. Formation of single
Wolf-Rayet stars probably took place as a result of such intense mass loss
(Bisnovatyi--Kogan, Nadyozhin, 1972). Outflowing atmospheres of luminous massive
stars are described
by equations of radiation hydrodynamics with radiative heat conductivity. The
flux goes through a critical point where the velocity is equal to the local
isothermal speed of the sound. The main difference between the outflowing
atmospheres in blue and evolved supergiants is a value of the optical depth (
average Rosseland ) in the critical point - $\tau_{cr}$. In blue supergiants we
have $\tau_{cr}<1$, and acceleration is essential only in parts of the
radiation spectrum around the spectral lines, but in evolved yellow
supergiants there is $\tau_{cr}>>1$ and the whole radiation flux takes part in
the flow acceleration.

Theory of the mass outflow from the very luminous evolved stars was developed
by Bisnovatyi-Kogan, Zeldovich, (1968), and self - consistent method for evolutional calculations
with a mass loss was proposed by (Bisnovatyi - Kogan, Nadyozhin, 1969,
Bisnovatyi - Kogan, Nadyozhin, 1972).  The method was based on finding the
solution of the outflowing envelope which continuously passes the isothermal
speed singular point and smoothly enters the second singular point at infinity
where $v\to v_{\infty}$, $T\to 0$. The shortcoming of the consideration of
~\cite{bkn}, where evolution with self - consistent mass loss was first
calculated, was ignorance of the fact, that the optical depth in the flow is
decreasing and far from the star $\tau\to 0$ there is no influence of matter
on the radiation.  Formally equations used in this paper described only optically
thick outflow and could not be extended to infinity.

The aim of this paper is to correct this shortcoming and to give a unified
description of the flow which is started at large optical depth and goes to
infinity at $\tau=0$. The approximate system of equations based on Eddington
approximation for the radiation is derived and solution is found which is
continuous in both singular points.

The  recipe based on the approximate
treatment of the outer boundary condition was proposed for the problem of the
mass loss by ~\cite{zyt},~\cite{zyt2}. Similarly to the case of a static
atmosphere it was suggested that at the stellar photosphere where
$T=T_{eff}=(L/4\pi r^2 \sigma)^{1/4}$ there is a fixed value of the optical
depth $\tau_{ph}=2/3$, or later $\tau_{ph}=8/3$ in the papers of
(Kato, 1985,  Kato, Iben, 1992).  The basic approximation made in these
papers was based on the approximate choice of the value of $\tau_{ph}$, which
instead of its expression $\tau=\int\limits_r^\infty \kappa\rho dr$ was taken
as $\tau_{ph}\simeq\tilde \tau_{ph}=(\kappa \rho r)_{ph}$.  It looks out here,
that we avoid all problems connected with the singular point at infinity,
because the region of solution is restricted by $r<r_{ph}$. In fact this
approximation creates another problems. In the case of incomplete ionization
and rapid change of the opacity as a function of $T$ and $\rho$ this
approximation could introduce rather big errors into the solution.
The method considered is free from such
shortcomings also. In the solution presented below we have $\tau_{ph}=4$
and $\tilde\tau_{ph}=3.75$. Recently another approximate approach for solution
of the
outflowing atmosphere was proposed by ~\cite{shar} Expanding layers of the star
are divided into subsonicaly extending photosphere where the stationary
momentum equation is adopted, and the wind where the accelerating up to
$v_{\infty}$ velocity of the outflow is approximated by the prescribed profile
with 3 parameters which are determined from matching and additional conditions.

\section{Thermodynamic relations in
outflowing envelope at arbitrary optical depth.}

In deep layers, at large optical depth $\tau\gg 1$ we have a usual equation
of state of a mixture of ideal gas with a black - body radiation.

\begin{equation}
\label{1a}
P(\rho, T)=\fr{aT^4}{3}+\rho{\cal R}T\mbox{.}
\end{equation}

\noindent
When $\tau$ is small it is not correct to  speak
about equilibrium between radiation and matter. Since local
thermodynamical equilibrium  is still presumed
we take isotropic radiation component together with a gas pressure in the form

\begin{equation}
\label{1b}
P^{isotr.}_{\tau\to 0}=\fr{aT^4}{3}\cdot\tau+\rho{\cal R}T \mbox{.}
\end{equation}

\noindent
Multiplier $\tau$ in (2) ensures zero decreasing contribution of a
$\fr{aT^4}{3}$
component at the infinity.

\noindent
An increasing anisotropy of
a pressure of radiation coming from the star
at decreasing $\tau$ is taken into account in Eddington approximation.
Basing on the solution of  the radiative transfer equation in Eddington approximation
for the spherically symmetric case with variable Eddington factor
(Sobolev, 1967), we introduce the following
approximate representation of thermodynamic relations for radiation

\begin{equation}
\label{1c}
P_r=\frac{aT^4}{3}\left(1-e^{-\tau}\right)+\frac{L(r)_{th}}{4\pi
r^2c}\mbox{,}
\end{equation}

\begin{equation}
\label{1d}
\rho E_r=aT^4\left(1-e^{-\tau}\right)+\frac{L(r)_{th}}{4\pi
r^2c}\mbox{.}
\end{equation}

\noindent
Term in (3), (4) containing $L_{th}$  is to be determined from
a solution of self-consistent set of equations . This
term essentially contributes only when $\tau$ is small.
This fact allows us to simplify further calculations by writing out
$L^{\infty}_{th}$ instead of $L_{th}(r)$.

\section{ Basic equations. }
A system of equations of radiation hydrodynamics describing continuous
transition between optically thick and optically thin regions for the
stationary outflow is written as:

\begin{equation}
\label{1}
u\frac{du}{dr}=-\frac{1}{\rho}\frac{dP_g}{dr}-\frac{GM(1-
\tilde L_{th})}{r^2},\\
\end{equation}

\begin{eqnarray}
\mbox{where}\quad\tilde L_{th}=\frac{L_{th}(r)}{L_{ed}}
\quad\mbox{,}\quad L_{ed}=\fr{4\pi cGM}{\kappa}\nonumber
\end{eqnarray}

\begin{eqnarray}
L&=&4\pi\mu\left(E+\frac{P}{\rho}-\frac{GM}{r}+\frac{u^2}{2}\right)+L_{th}(r)\\
L_{th}&=&-\frac{4\pi r^2c}{\kappa\rho}\left(\frac{dP_r}{dr}-\frac{E_r\rho -
3P_r}{r}\right)\\
\fr{\dot M}{4\pi}&\equiv\mu&=\rho ur^2\\
P&=&\frac{aT^4}{3}\left(1-e^{-\tau}\right)+\frac{L^{\infty}_{th}}{4\pi
r^2c}+P_g\\
E\rho&=&aT^4\left(1-e^{-\tau}\right)+\frac{L^{\infty}_{th}}{4\pi r^2c}+E_g \\
P_g&=&\rho{\cal R}T\\
E_g&=&\frac32 {\cal R}T \\
\tau&=&\int\limits_r^\infty \kappa\rho dr
\end{eqnarray}

\noindent Where $L$ - is a constant total energy flux consisting of the
radiative energy transfer together with the energy of the matter current, $u$
is a rate of the outflow, $\kappa$ is an opacity, assumed to be constant, $a$ is the constant
of a radiative energy density, $\cal R$ is a gas constant.

\noindent
We consider here the flow in the gravitational field of constant mass $M$,
neglecting self-gravity of the outflowing envelope outside the critical point
$m_{en}$. This approximation is very good in a realistic case of $m_{en}<<M$
(Zytkow, 1973).
This system of equations provides  a  description of
a stationary outflowing envelope accelerated by a radiative force
at arbitrary optical depth where continuum opacity is prevailed.
In optically thick limit $\tau\to\infty$,
when terms in (9) , (10) with $L^{\infty}_{th}$ are negligible,
and $E_r\rho=3P_r$
a solution of this system was obtained by ~\cite{bkpaper}.
In case of small $\tau$ from (9), (10) for the anisotropic radiation
flux we get: $E_r\rho\simeq P_r$, what follows from transfer equation in
Eddington approximation. (Sobolev, 1967).

When optical depth becomes small separation of radiation and matter should
be taken into consideration. It means that
only a part of radiation is interacting with outflowing gas envelope.
For this part of quantums we still assume LTE to be valid, what means that we
take the temperature in (3) - (4) to be the same as in (11) - (12). For the rest
part of radiation that is not interacting with outflowing matter, another
'temperature'- mean energy of quantums should be introduced. This part of
radiation transfers momentum to the outflowing matter and thus produces only
the anisotropic part of the pressure, that is of the order of $L_{th}/r^2$.  We
avoid the problem of treating this another 'temperature' by solving the
radiative transfer equation in momentum form (7).  It is solved together with
the equation of motion (5), energy conservation (6) and thermodynamical
relations for the outflowing gaseous matter and radiation (9) - (10).

Equation of motion (5) is written in the form where the effect of the pressure
gradient is explicitly written only for the gas component and all
effects of radiation is embedded into the last term. This term is obtained
from
the transfer equation (7) when taking into account relations (9), (10).
In case of large optical depths this  approach insures that acceleration
is providing by the total (isotropic = radiation + gas ) pressure gradient,
and in case of small $\tau$ by gas pressure gradient and pressure of radiation
flux. In intermediate region this terms will compete in accordance with (3),
(4) and (7).
About the difference of this system of equations from previously used see the
Discussion.

Substituting (7) - (10) into  (6) we get

\begin{eqnarray}
\frac{L}{4\pi}=\rho u
r^2\left(E+\frac{P}{\rho}-\frac{GM}{r}+\frac{\mu^2}{2\rho^2 r^4}\right)\nonumber
\end{eqnarray}

\begin{equation}
\label{first0}
-\frac{r^2c}{\kappa\rho}\left\{\frac{d}{dr}\left[\frac{(1-e^{-\tau})
aT^4}{3}+ \frac{L^{\infty}_{th}}{4\pi r^2c}\right]+2\frac{L^{\infty}_{th}}{4\pi
r^3c}\right\} \mbox{.}\\
\end{equation}

\noindent
Differentiating in ~(\ref{first0}) with taking into account that
$d\tau=-\kappa\rho dr$ will result in a first differential equation

\begin{eqnarray}
\lambda r^2(1-e^{-\tau})\frac{dT}{dr}=\mu\left\{\frac52{\cal R}T-
\frac{GM}{r}+\frac{\mu^2}{2\rho^2r^4}\right\}\nonumber
\end{eqnarray}

\begin{eqnarray}
-\frac{L}{4\pi}+\mu\left[2\frac{L^{\infty}_{th}}{4\pi r^2c\rho}
+\frac{4}{3\rho}(1-e^{-\tau}) aT^4\right]\nonumber
\end{eqnarray}

\begin{equation}
\label{first}
+\fr13ar^2cT^4\ex \mbox{.}
\end{equation}

Here we introduce a coefficient of a heat conductivity

\begin{equation}
\label{lmbd}
\lambda=\fr {4acT^3}{3\kappa\rho}\mbox{.}
\end{equation}

In a limiting case of $\tau \to \infty$ this equation coincides with a
corresponding  equation from ~\cite{bkpaper}, when taking into account that

\begin{eqnarray}
\fr{L_{th}}{4\pi r^2}\ll caT^4.\nonumber
\end{eqnarray}

\noindent
Using (11) , (8) in (5) will get

\begin{equation}
\label{f1}
\left(\fr{{\cal R}T}{\rho}-\fr{\mu^2}{\rho^3 r^4}\right)\fr{d\rho}{dr}=
-{\cal R}\fr{dT}{dr}+\mu^2\fr{2}{\rho^2 r^5}-\fr{GM}{r^2}(1-\tilde L_{th})\mbox{.}
\end{equation}
\noindent

The system ~(\ref{first}) , ~(\ref{f1}) need to be completed from (13)
by relation

\begin{equation}
\label{dpth1}
\fr{d\tau}{dr}=-\kappa\rho\mbox{.}
\end{equation}

\noindent
The equation ~(\ref{f1})  has a singular point, where
the left hand side of it vanishes

\begin{equation}
\label{border1}
\fr{{\cal R}T_{cr}}{\rho_{cr}}=\fr{\mu^2}{\rho^3_{cr} r^4_{cr}}\mbox{.}
\end{equation}

This point corresponds to the "isothermal
sonic" point where

\begin{eqnarray}
u^2=u^2_s\equiv\left(\fr{\partial P}{\partial \rho}\right)_T\mbox{.}\nonumber
\end{eqnarray}

\noindent
Using ~(\ref{lmbd}) ,  (7) gives
\begin{equation}
\label{Lth}
L_{th}=-4\pi r^2\left[\lambda\fr{dT}{dr}(1-\ex)-\fr13 acT^4\ex\right]\mbox{.}
\end{equation}

\noindent
The second singular point of ~(\ref{first}) - ~(\ref{dpth1}) is situated
at infinity, where

\begin{equation}
\label{border2}
T=0,\quad\rho\sim\fr{1}{r^2}\to 0,\quad u\to const=u_{\infty},\quad\mbox{when}\quad
r\to\infty\mbox{.}
\end{equation}

\noindent
This condition is related to the fact that far from the star the density in
stellar wind is very small. In reality the wind may be treated as stationary
only up to the limiting radius $r_{lim}\simeq v_{\infty}t>>r_{cr}$, where $t$
is the characteristic mass-loss time of the star. So the formal solution with
outer boundary condition (21) is very close to the real solution with
conditions at $r_{lim}$.
This approximation is a common way to consider a well-developed solar wind
problem (Parker, 1963).
Let us introduce nondimensional variables

\begin{equation}
\label{dim1}
\tilde T(r)=\fr{T(r)}{T_{cr}}\quad,\quad\tilde
\rho(r)=\fr{\rho(r)}{\rho_{cr}}\quad,\quad\tilde
L_{th}=\fr{L_{th}}{L_{ed}}\mbox{,}
\end{equation}

\begin{eqnarray}
\tilde x=\fr{r_{cr}}{r}\mbox{.}\nonumber
\end{eqnarray}

\noindent
After transformations we obtain dimensionless system of equations

\begin{eqnarray}
\fr{d\rho}{dx}=\left(\fr{x^4}{\rho^3}-\fr{T}{\rho}\right)^{-1}
\biggl\{\fr{dT}{dx}\left(1+A_1(1-\ex)\fr{T^3}{\rho}\right)
\end{eqnarray}

\begin{eqnarray}
-A_3+\fr{1}{4}\fr{A_1\ex}{A_5}\fr{T^4}{x^2}+
2\fr{x^3}{\rho^2}\biggr\}\quad\mbox{,}\nonumber
\end{eqnarray}

\begin{eqnarray}
\fr{dT}{dx}=-\biggl(\bern
\end{eqnarray}

\begin{eqnarray}
+\fr{\ex}{4A_2A_5}\fr{T^4}{x^2}+2L^{\infty}A_3A_5\fr{x^2}{\rho}-
\fr{A_4}{A_2}\biggr)\fr{A_2\rho}{T^3(1-\ex)}\quad\mbox{,}\nonumber
\end{eqnarray}

\begin{equation}
\fr{d\tau}{dx}=\fr{\rho}{A_5x^2}\quad\mbox{.}
\end{equation}

\noindent
Where $L^{\infty}\equiv\tilde L^{\infty}_{th}$.
To simplify writing here and further we omit tilde.
Dimensionless coefficients $A_i$ are:
\begin{equation}
\label{dimcoef1}
A_1=\fr{4aT_{cr}^3}{3\rho_{cr}\cal R}\quad,\quad
A_2=\fr{3\kappa\mu}{4ac}\fr{\rho_{cr} {\cal R}}{r_{cr} T^3_{cr}} \mbox{,}\\
\end{equation}

\begin{equation}
\label{dimcoef2}
A_3=\fr{GM}{r_{cr}{\cal R}T_{cr}}\quad,\quad
A_4=\fr{3\kappa\L}{16ac\pi}\fr{\rho_{cr}}{r_{cr}T_{cr}^4}\mbox{.}
\end{equation}

\noindent
Physical sense of $A_i$ parameters have been revealed in ~\cite{bkpaper}.
Additional parameter is

\begin{equation}
\label{dimcoefA5}
A_5=\fr{1}{r_{cr}\kappa\rho_{cr}}\quad,
\end{equation}
\noindent
the reciprocal $1/A_5$ is of the order of optical depth in the critical point.

\noindent
Condition of transition of the solution through  the critical
point confines the number of dimensionless parameters. Equating
to zero expression in the figure brackets of (23) in the critical
point with account of (24) will get

\begin{eqnarray}
A_4=\bigl(4 A_3 A_5 (1 - \ec) +8 A_5(-1 +\ec)\nonumber \\
+ 4A_1^2 A_2 A_5\bigl(1 - 2\ec + (\ec)^2\bigl)\\
+A_2 \bigl(12 A_5 + A_3(-4 A_5 + 8 A_5^2 L^{\infty})\bigl)\nonumber \\
+A_1 A_2 \bigl(16A_5(1 - \ec)\nonumber \\
+A_3(4A_5 (-1 + \ec) + A_5^2 (8 L^{\infty}\nonumber \\
- 8 \ec L^{\infty})\bigl)\bigl)+\ec)/4\bigl ( A_5 +
A_1 A_5 (1-\ec) \bigl )\mbox{.}\nonumber
\end{eqnarray}

\noindent
All dimensional parameters of the flux: $T$ , $\rho$ , $r$ could be expressed
as a function of dimensionless parameters $A_i$ and a dimensional combination
of physical constants ~\cite{bkpaper}

\begin{eqnarray}
r=(\fr{4a\kappa}{3c})^{2/5}\fr{(GM)^{7/5}}{{\cal R}^{8/5}}\fr{1}{(A_1^2 A_2
A_3)^ {2/5} A_3 x}\mbox{,}\nonumber
\end{eqnarray}

\begin{eqnarray}
\rho=(\frac{3{\cal R}}{4a})^{1/5}\left(\fr{c{\cal R}^{1/2}}{\kappa GM}\right)^{6/5}(A_1^2 A_2
A_3)^{6/5} \fr{\tilde \rho}{A_1}\mbox{.}\nonumber
\end{eqnarray}

\noindent
In numerical calculations coefficient $A_5$ should be expressed as a function
of $A_1$ - $A_3$ and
a nondimensional combination of physical constants and thus it is not an
independent parameter

\begin{equation}
\label{A5}
A_5=\left(\fr34 \right)^{1/5}\fr{A_3^{1/5}{\cal R}^{4/5}}{A_1^{3/5}
A_2^{4/5} k^{1/5} a^{1/5} c^{4/5} (GM)^{1/5}}\mbox{.}
\end{equation}

\section {Numerical solution }

In order to satisfy boundary conditions far from the star we need to integrate
(23) - (25) from the critical point outward to the infinity.  We exit the
critical point by means of expansion formula.  Expanding the solution in
critical point $x=T=\rho=1$ in
powers of $(1-x)$  we have
\begin{equation} \label{appr1}
T=1+a(1-x)\mbox{,}\\ \end{equation}

\begin{equation}
\label{appr2}
\rho=1+b(1-x)\mbox{.}
\end{equation}

\noindent
Similarly
\begin{eqnarray}
e^{-\tau}\simeq\ec (1+\fr{ y }{A_5})\mbox{,}\nonumber
\end{eqnarray}

where $y=1-x$.

For the $a$ and $b$ coefficients we get

  \begin{eqnarray}
  b=(-12 A_2 A_5-4 A_1 A_2 A_5+4 A_2 A_3 A_5+4 A_4 A_5 \nonumber\\
 + 4 A_1 A_2 A_5 \ec-8 A_2 A_3 A_5^2 L^{\infty}\\
  -\ec )/\left(4 A_5 (-1 + \ec)\right) \mbox{,}\nonumber\\
  a=(-c_1-(c_1^2-4 c_2 c_0)^{1/2})/(2 c_2)\mbox{.}
  \end{eqnarray}

The coefficients $c_i$ due to their complicated form are adduced in
Appendix.

Numerically integrating we escape the critical point
by means of the expansion formulas ~(\ref{appr1}), ~(\ref{appr2}) . Then
integrating outward to the infinity we
satisfy the boundary conditions ~(\ref{border2}) .  Results of the numerical
calculations are shown in Fig.1-2. Curves at this figures correspond to the
following dimensional values
of the dimensionless parameters:
$A_1=50$, $A_2=10^{-4}$, $A_3=43.88$, $\tau_{cr}=125$, $L^{\infty}_{th}=0.6$.
This set of parameters corresponds to the following values at critical
point: $T_{cr}=1.4\cdot 10^4 K$, $r_{cr}=2.6\cdot
10^{13} {\rm cm}$, $\rho_{cr}=6.6\cdot 10^{-12} g/{\rm cm}^3$.

The behavior of the solution with rapidly
decreasing Mach number at $r<r_{cr}$ shows, that it may be matched
to the static solution for the core. In reality the opacity peak is situated
near critical point, opacity inside is decreasing and the inside velocity drop
is more rapid, then in the case of $\kappa=const$ (Bisnovatyi - Kogan and
Nadyozhin, 1972).
The effective temperature of the photosphere
may be obtained from: $L^{\infty}_{th}/4\pi r^2=\sigma T^4$.
For the given set of parameters we get: $x_{ph}=0.03$,
$\tau_{ph}=4$,
$T_{eff.}=0.06$, and $\tilde\tau_{ph}=3.75$.
Conditions in critical points impose two relations
upon the set of nondimensional parameters. As it was shown the solution
depends upon the following nodimensional parameters: $A_1$, $A_2$, $A_3$, $A_4$,
$\tau_{cr}$ , $L^{\infty}_{th}$. When numerically treated $A_4$ was expressed
as a function
of the remaining dimensionless parameters when taking into account that
in sonic point $x=\rho=T=1$ ~(\ref{dim1}). Parameter $A_3$ is to be numerically
determined from the condition at the infinity where $T=0$ ~(\ref{border2}).

When $r\to\infty$ velocity tends to constant and thus $\rho\sim 1/r^2$.
Only the unique value of $\tau_{cr}$ allows to obtain the proper behavior of
$u$ (or $\rho$) at the infinity.
Varying $\tau_{cr}$, we adjust the behavior of $u$
to get $u\to u_{\infty}$ at $r\to\infty$.

At $r=\infty$ the solution could be represented in an expansion form .
In this case $L^{\infty}_{th}$ could be directly determined from
~(\ref{Lth}), and the appropriate expansion.

At the stellar core optical depth may be taken arbitrary large, and it
is necessary to match only $T$ and $\rho$.
Above mentioned method of obtaining the unique value of $\tau_{cr}$ allows to
avoid the problem of fitting $\tau$ when integrating to the stellar
core.

All other dimensionless parameters of the envelope and parameters of the static
core could
be obtained only when matching the
solution for the stellar core with the solution for the expanding layers.
In this approach all the treatment will be
fully self - consistent.

\noindent
In our
treatment we have not used the second expansion and thus could not specify
uniquely the $L^{\infty}_{th}$.

Our aim was to develop a method of a calculating the
parameters of the flux in arbitrary $\tau$. Fully self - consistent treatment
which may be applied to the real star is under the consideration.

\section{Discussion}

Solution for the spherically-symmetrical
stationary outflowing envelope accelerated by the radiative force
in arbitrary optical depth case was obtained in this work.

We have introduced thermodynamical relations for the matter flux
with partially separated radiation. This approach provides
satisfactory description of the problem for the arbitrary $\tau$.
In case of $\tau\to\infty$ equation of state yields
$E_r\rho=3P_r$ and our method reproduces results of ~\cite{bkpaper}. Also when
$\tau\to 0$: $E_r\rho\simeq P_r$ accords the result
obtained from the radiative transfer equation in Eddington approximation.

As  a result of this treatment we have introduced a system
of differential equations. This system provides a continuous transition
of the solution between optically thin and optically thick regions.
To satisfy boundary conditions the solution should proceed through the
critical point where the speed of the flux equals the local isothermal sound
speed. We have derived analytically approximate representation of the solution
at the vicinity of the sonic point. Using this representation we numerically
integrate the system of equation from the critical point to satisfy conditions
at the infinity.

Beginning from (Zhytkow, 1973), the obtaining of outer boundary conditions
was oversimplified, making a wrong impression, that regions
with $\tau < 1$ , have no influence on the mass outflow.
Zhytkow imposed boundary conditions at the photosphere.
That was also the result of that there was no method for the self-consistent
description of radiation and matter at small $\tau$.
It should be mentioned that most of the papers only inherited the method
developed by Zhytkow.
As soon as it is possible to describe correctly the whole region of the flow
one should forthwith impose correct boundary conditions.
Problem of the infinite boundary conditions was elaborated for the
theory of Solar wind.

In papers of Zytkow, it was made an attempt to describe the whole flow. The
description
was based on Paczynsky approximation (Paczynski, 1969), that rouphly takes
into account
the dilution of the radiation flux.
For $\tau<2/3$ for the part of the equation of motion, that describes
pressure, it was taken:

$$\fr{d\rho}{dx}\sim-\fr{\partial P_g}{\partial T}
\fr{dT}{dr}+\fr{\kappa\rho}{4\pi c}\fr{L}{r^2}\mbox{,}\qquad
{\rm where}\qquad
\fr{dT}{dr}=-\fr{L}{4\pi r^2\lambda}-\fr{1}{2} f(\tau) T_0
R_0^{1/2}r^{-3/2}\mbox{,}$$ where $T_0^4=L/4\pi ac R_0^2$ - the effective
temperature of the photosphere, and $L$ is constant for $\tau<2/3$,
$f=1-(3/2)\tau$.  The second term in the relation for the temperature gradient
originated from the approximation of regions of small $\tau$ ( with Eddington
factor 1 ), with regions of $\tau>>1$ ( with factor 1/3 ) and is to describe
roughly the process of the radiation dilution.  The shortcoming descended from
the description of Zhytkow concerns this last effective surface ( with
$T=T_{eff}$, $\tau =2/3$ ).  The 'temperature' of free, not interacting
quantums corresponds to the prescribed $T_{eff}$ of the photosphere. It can be
not a bad approximation, if mostly neglecting the influence of the regions with
small $\tau$, but to describe the whole flow, this approach seems to be too
rude.  It seems obvious that the solution is artificially restricted by taking
this prescribed separation with the certain value of the effective temperature
of the photosphere.

When $\tau$ becomes
small, separation of the radiation and matter progresses in two parts (two
'temperatures', see appropriate discussion), with accordance to the
transfer equation in momentum form (7).
If not treating the transfer equation, together with equations of hydrodynamics,
the energy conservation will be broken for the radiation at the regions of
small $\tau$.  That will only enlarge the uncertainty within the extended
atmosphere.

The main difference of our approach with previously published is that
we use the equation of motion, in which the effects of the gas pressure are
separated from the effects of the radiation, together with the transfer
equation (7). The effect of the gas pressure
gradient
that is valid at small $\tau$ as well as in deep interior is
explicitly written. At $\tau>>1$ the main driving force originates from
the gradient of the pressure of the gas together with the radiation.
Far from the star acceleration occurs mainly due to the momentum transfer
from the radiation to matter. In the way the equation (5) is written
it is correct for arbitrary optical depth, if, of course, to calculate
$L_{th}$ from (7).
This description provides the proper competition of this essentially different
effects of the radiation.
Contrary to other authors which used approximations
obtained from (7), we are resulting from the 'equations of state' (9) - (10).
We believe, that this approach is almost free from the shortcomings mentioned
above.

We have used equation of state which corresponds to constant $R_g$.
Account of variable $R_g$, $\dot M$ and $\kappa$ are of principal importance
for the real stars as it was described by (Bisnovatyi-Kogan, Nadyozin, 1972).
This complication does not change our method and qualitative results.  When
one will take into account lines effects, the importance to describe correctly
the region of small $\tau$ will forthwith be grater.

As it was mentioned at the beginning, the problem of taking $\tilde\tau$
instead of $\tau$ needs to be treated more carefully.
In this paper we simply assumed $\kappa$ to be constant.
When considered far from the star velocity of the flux may be
approximately taken constant, and thus from $\dot M=4\pi\rho u r^2$,
get $\rho\sim 1/r^2$. In this case $\tau\simeq\tilde\tau$.
Otherwise if we consider a power law of
$\kappa=\rho^{\alpha}$ it is easy to obtain
$\tau_{ph}=(\kappa\rho r)_{ph}/(1+2\alpha)$, what is $(1+2\alpha)$ times
smaller than $\tilde\tau_{ph}$.
For steeper (exponential) decrease of $\kappa$ with $r$ the difference
between $\tau_{ph}$ and $\tilde\tau_{ph}$ may be much larger.

In real stars opacity may increase more steep. Even if
temperature is decreasing smoothly, partial recombination of
ions will cause an increase of $\kappa$ ( Cox, Tabor 1975 ).
Hence it will create regions where flow is accelerated and
$u$ can not be taken constant and thus law for $\rho$ will
be far from $1/r^2$. In this case rather big errors could be introduced
when taken $\tilde\tau$ instead of $\tau$.

On a supergiant phase massive stars may have regions of not
fully ionized $\rm H$ and $\rm He$ what cause an increase of the opacity
which leads to the acceleration in continuum ( Bisnovatyi - Kogan, Nadyozhin
1969, 1972 ).

To take into account real physical effects our method
should be improved by taking non-constant $\kappa$ and variable ionization
degree. If improved this way theoretical approach presented in this work should be
useful for self - consistent simulations of the massive stars
evolution with mass loss.

{\small This work was partly supported by RFBR
grant 96-02-17231, grant 96-02-16553, and CRDF grant RP1-173,
Astronomical Program 1.2.6.5 .}

\onecolumn
\section{Appendix}
Further sophisticated calculations were made using
Mathematica 2.2 .
$$
\begin{array}{rcl}
c_0&=&\bigl(A_1 A_2^2 A_3^2 A_5^4 96(-1 + \ec) (L^{\infty})^2+
A_5^3 (-40 A_2^2 A_3 L^{\infty}+A_1^2 A_2^2 A_3 (-160 + 320 \ec -\\
& & \\
& & 160 (\ec)^2) L^{\infty}+A_2 (A_3 176 (1 - \ec) L^{\infty} +
A_3^2 (-48 + 48 \ec) L^{\infty})+\\
& &\\
& &A_1 (A_2 A_3 (80 + 96 A_4 + (-160 - 96 A_4) \ec + 80 (\ec)^2)
       L^{\infty} +
A_2^2 (A_3^2 96 (1 - \ec) L^{\infty} + A_3 392 (-1 + \ec)
       L^{\infty}))) -\\
& & \\
& & 2 (\ec) ^2 + A_5 (2 (\ec)^2  + \ec (-2+ 8 A_4 ) +
A_2 \ec (-29 + 8 A_3 ) + 2A_3 (1
    \ec  - (\ec)^2 ) +\\
& &\\
& &2A_1^2 A_2\ec (1 - 2 (\ec)  + (\ec)^2 ) +
 A_1(-2A_4\ec  + 2A_4(\ec)^2  +
A_2 (-7 \ec  + 7 (\ec)^2  + A_3 (-2 \ec  + 2\\
&&\\
&& (\ec)^2 )))) +
A_5^2 (-96 + A_2^2 (-60 + 20 A_3) - 72 A_4 + (192 + 72 A_4) \ec -
96 (\ec)^2 + A_3 (24 + 24 A_4 +\\
& & \\
& &(-48 - 24 A_4) \ec + 24 (\ec)^2) + A_1^3 A_2^2 (-56 + 168 \ec -
168 (\ec)^2 + 56 (\ec)^3) +\\
& & \\
& &A_1^2 (A_2 (24 + 80 A_4 + (-72 - 160 A_4) \ec +
(72 + 80 A_4) (\ec)^2 - 24 (\ec)^3) +A_2^2 (-292 + 584 \ec - 292 (\ec)^2 +\\
& & \\
& &A_3 (80 - 160 \ec + 80 (\ec)^2))) +
A_2 (232 + 20 A_4 + A_3^2 (24 - 24 \ec) - 232 \ec +\\
& & \\
& &A_3 (-152 + 152 \ec - 16 \ec L^{\infty} )) +
A_1 (-24 A_4 - 24 A_4^2 + (48 A_4 + 24 A_4^2) \ec - 24 A_4 (\ec)^2 +\\
& & \\
& &A_2^2 (-392 + A_3 (196 - 196 \ec) + 392 \ec +
A_3^2 (-24 + 24 \ec)) +
A_2 (160 + 196 A_4 + (-320 - 196 A_4) \ec + \\
&&\\
& & 160 (\ec)^2 +A_3 (-56 - 48 A_4 + (112 + 48 A_4) \ec - 56 (\ec)^2 +
L^{\infty} (4 \ec  - 4 (\ec)^2 )))))\bigl)/\bigl(2 (-1 + \ec)\bigl)\mbox{,}\\
\end{array}
$$
$$
\begin{array}{rcl}
c_1&=&A_5^2 (8 + A_2 (56 - 20 A_3) - 20 A_4 + A_3 (12 - 12 \ec) - 8 \ec +\\
&&\\
&&A_1^2 A_2 (8 - 16 \ec + 8 (\ec)^2) +\\
&&\\
&&A_1 (-12 A_4 + 12 A_4 \ec + A_2 (48 - 48 \ec + A_3 (-12 + 12 \ec)))) +\\
&&\\
&&A_5^3 (32 A_2 A_3 L^{\infty} + A_1 A_2 A_3 (16 - 16 \ec) L^{\infty}) + 5 A_5 \ec\mbox{,}
\end{array}
$$

$$
\begin{array}{rcl}
c_2&=&A_5^2\left(8 - 8 \ec \right)\mbox{.}
\end{array}
$$

\newpage
\section{Firure captions}

{\bf Fig.1}
Results of the integration of equations (23) - (25) from the critical point $x=1$
outward to the infinity $x=0$.
Points $x=1$ and $x=0$ are the singular points of our system.
A solution shown for  $A_1=50$, $A_2=10^{-4}$, $A_3=43.89$, $\tau_{cr}=125$,
$L^{\infty}_{th}=0.6$,
passes through the critical point ($T_{cr}=1.4\cdot 10^4 K$, $r_{cr}=2.6\cdot
10^{13} {\rm cm}$, $\rho_{cr}=6.6\cdot 10^{-12} g/{\rm cm}^3$) and satisfies 
boundary conditions ~(\ref{border2}).

\noindent
{\bf Fig.2} Dimensionless velocity (solid line) and $\mbox{Mach number}=
u\cdot\left( \fr{\partial P}{\partial \rho} \right)_s^{-\fr12} $ (dashed line).
Inwardly decreasing Mach number ensures matching the static core
even in the
case of $\kappa=const$. When $r\to\infty$ velocity tends to constant
$v_{\infty} \simeq 16 km/s$ and $\rho\sim 1/r^2$ . Solution is shown for the
same values of nondimentional parameters as on
Fig. 1.

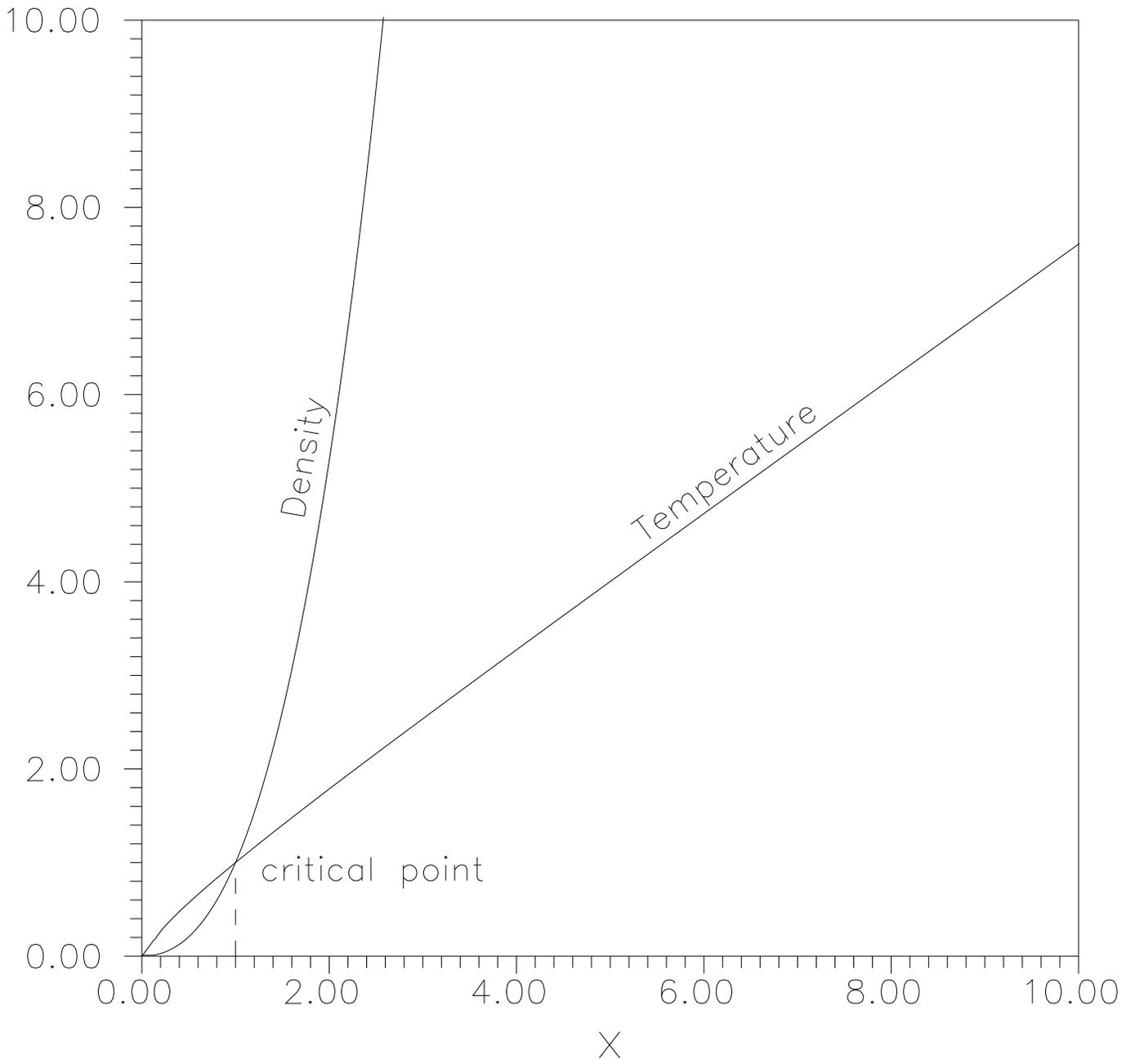
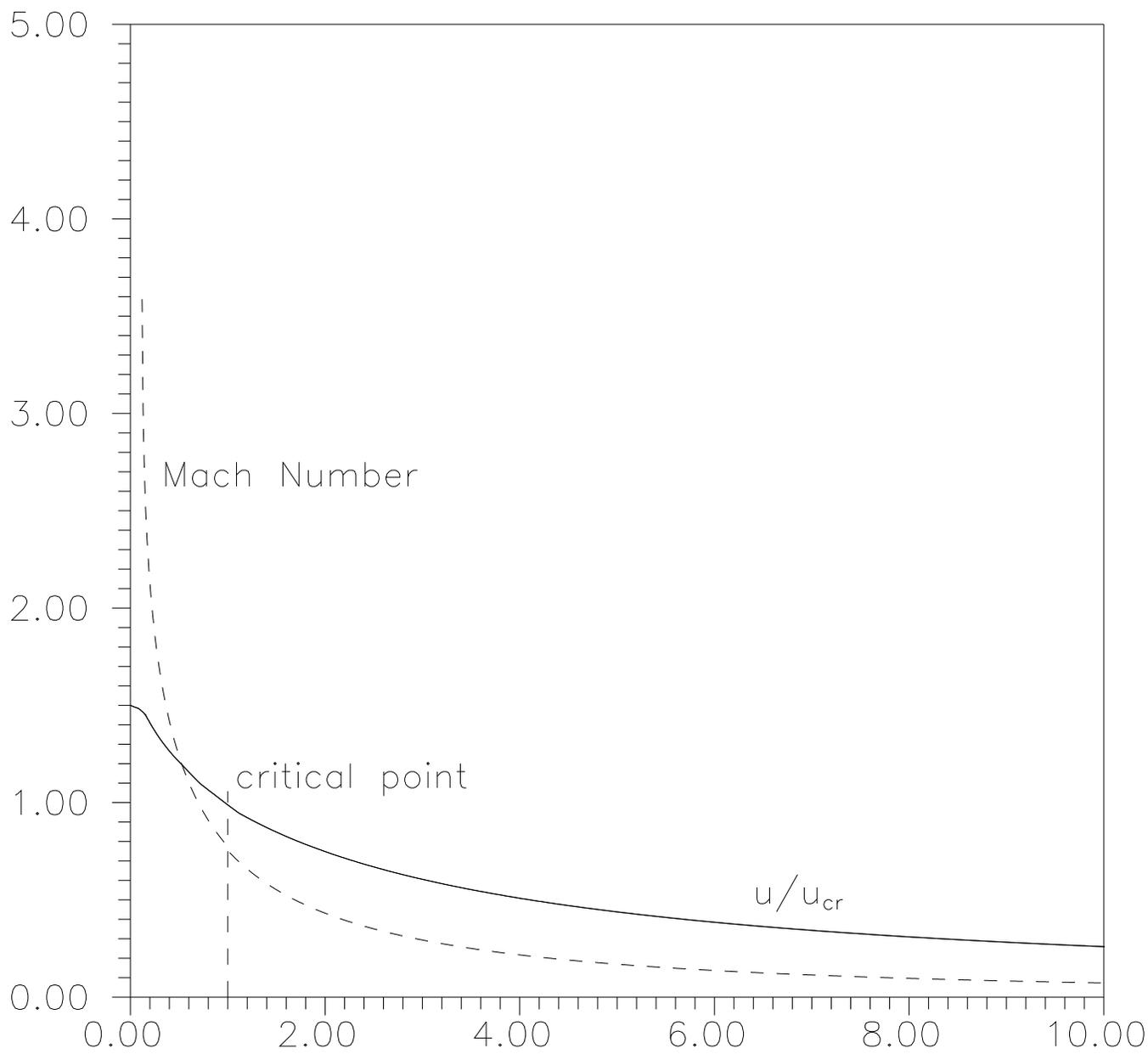
\end{document}